\documentclass[sigconf]{acmart}

\usepackage{subcaption}
\usepackage{caption}

\AtBeginDocument{%
  \providecommand\BibTeX{{%
    \normalfont B\kern-0.5em{\scshape i\kern-0.25em b}\kern-0.8em\TeX}}}

\copyrightyear{2021}
\acmYear{2021}
\setcopyright{rightsretained}
\acmConference[CHI '21]{CHI Conference on Human Factors in Computing Systems}{May 8--13, 2021}{Yokohama, Japan}
\acmBooktitle{CHI Conference on Human Factors in Computing Systems (CHI '21), May 8--13, 2021, Yokohama, Japan}\acmDOI{10.1145/3411764.3445309}
\acmISBN{978-1-4503-8096-6/21/05}

\begin{document}

\title[Covert Embodied Choice]{Covert Embodied Choice: Decision-Making and the Limits of Privacy Under Biometric Surveillance}





\author{Jeremy Gordon}
\affiliation{UC Berkeley}
\email{jrgordon@berkeley.edu}
\orcid{0000-0002-4514-5284}

\author{Max Curran}  
\affiliation{UC Berkeley}
\email{mtcurran@berkeley.edu}
\orcid{0000-0000-0000-0000}

\author{John Chuang}  
\affiliation{UC Berkeley}
\email{john.chuang@berkeley.edu}
\orcid{0000-0000-0000-0000}

\author{Coye Cheshire}  
\affiliation{UC Berkeley}
\email{coye@berkeley.edu}
\orcid{0000-0000-0000-0000}

\renewcommand{\shortauthors}{Gordon et al.}

\begin{abstract}
    Algorithms engineered to leverage rich behavioral and biometric data to predict individual attributes and actions continue to permeate public and private life. A fundamental risk may emerge from misconceptions about the sensitivity of such data, as well as the agency of individuals to protect their privacy when fine-grained (and possibly involuntary) behavior is tracked. In this work, we examine how individuals adjust their behavior when incentivized to avoid the algorithmic prediction of their intent. We present results from a virtual reality task in which gaze, movement, and other physiological signals are tracked. Participants are asked to decide which card to select without an algorithmic adversary anticipating their choice. We find that while participants use a variety of strategies, data collected remains highly predictive of choice (80\% accuracy). Additionally, a significant portion of participants became more predictable despite efforts to obfuscate, possibly indicating mistaken priors about the dynamics of algorithmic prediction.
\end{abstract}



\begin{teaserfigure}
  \includegraphics[width=\textwidth]{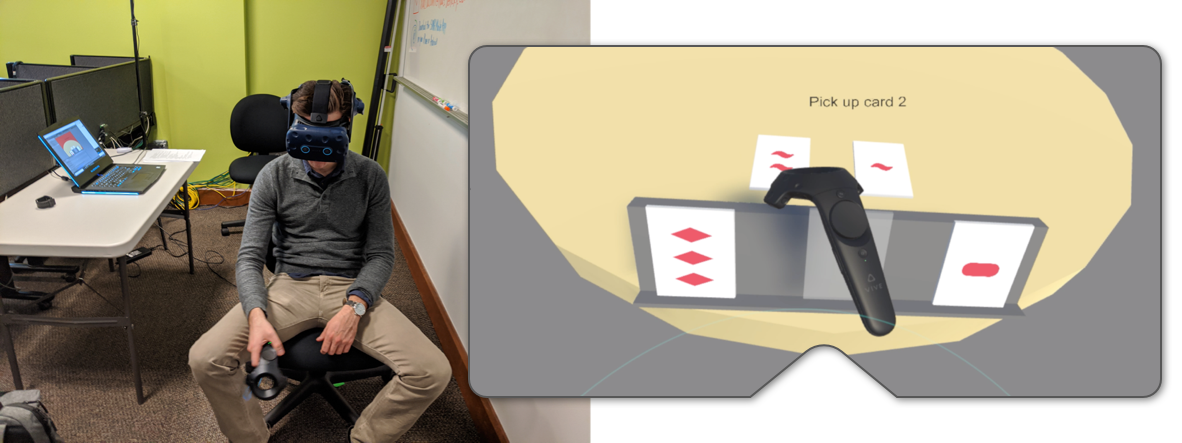}
  \caption{Left: participant in VR experiment. Right: participant field of view during card selection.}
  \Description{Left: participant sits on a chair holding VR controller and wearing VR head mounted display. Right: participant's field of view shows controller holding a card and bringing into their card holder.}
  \label{fig:teaser}
\end{teaserfigure}

\maketitle


\section{Introduction}

In a world in which sensing devices permeate both public and private domains, consumers~\cite{kirkup2000video}, renters~\cite{gagliordiAmazonIntrosAlexa}, voters~\cite{chaum2004secret}, and decision-makers of all kinds are now subject to unprecedented levels of surveillance during day-to-day life: from the tracking of search queries to face and `emotion recognition'~\cite{bullington2005affective}, to geospatial location and physiology~\cite{ball2010workplace}. The data produced by sensors and software capturing these signals are now and will continue to be analyzed by increasingly advanced statistical techniques and algorithms with a specific interest in inferring internal states that might predict the future actions of individuals. Constituting what Zuboff terms the `Big Other', this constellation of tracking infrastructure uses opaque mechanisms to detect and control behavior, and ultimately poses a threat to both individual privacy and democratic norms at large~\cite{zuboff2015big}. 

According to Nissembaum's contextual integrity heuristic, privacy is best understood as a set of appropriate information flows subject to contextual norms dependent on parameters such as the data subject, parties, information type, and principles of transmission~\cite{nissenbaum2009privacy}. As proposed by Sedenberg et al.\ 2017, the privacy threats posed by remote biometric sensing can be seen as shifts or breaches of these contextual norms: ``...when biosensed data like emotions or internal physiological states are systematically recorded and analyzed, all signals become magnified beyond their original natural public scope''~\cite{sedenberg2017window}. Indeed, remote sensing systems of this kind have already generated real-world contexts within which actors may hope to predict internal beliefs or preferences from biometric data, and for which the obfuscation of these beliefs may be beneficial to the individual. One such example is individually targeted price discrimination in a physical retail setting where video feeds may afford gaze tracking (see commercial use of "Smart-shelf" technology reported in 2013\cite{boulton2013snackmaker}).

Thanks to journalistic and public service efforts, an awareness of these threats is growing, but individuals still hold significant misconceptions about the sensitivity of information that can be inferred from these types of data~\cite{merrill2019sensing}, and what agency they have in protecting their privacy. 

In this work, we designed a Virtual Reality (VR)-based behavioral experiment in which participants completed an iterative binary decision task. In later trials, participants were informed of an adversary that is tracking their behavior and attempting to predict each choice. Questionnaire responses, post-task interviews and quantitative analyses of biometric data indicate that participants used a variety of behavioral strategies and felt confident in their agency in avoiding prediction. While some participants adjusted their behavior in ways that reduced prediction accuracy, a strong majority of trials could still be successfully classified based on biometric data despite these efforts.

\subsection{Background}

\subsubsection{Embodied Cognition and Decisions}

A valuable perspective on the privacy of decisions in a tracked environment comes from a multidisciplinary literature which integrates embodied, enactive, extended, and embedded accounts of cognition (referred to as `4E Cognition'). In a 2015 paper, Lepora and Pezzulo introduced the Embodied Choice framework (EC) which recasts decision-making as a dynamic and inherently active process rather than a sequential perception-decision-execution cycle~\cite{leporaEmbodiedChoiceHow2015}. In a visual decision task, the authors tracked the correlates of active consideration via movements of a mouse towards a target. These ideas are inspired by and consistent with Active Inference, a formal framework modeling situated action in which agents minimize their prediction error by entraining available sensors in such a way as to disambiguate competing perceptual hypotheses~\cite{seth2014cybernetic}. In this framework, percepts can be seen as hypotheses (e.g. \emph{the dog is hungry}, or \emph{the customer is planning to buy some flowers}), and the ocular motor outputs that produce saccades (eye movements) are seen as experiments aiming to confirm or deny prior beliefs~\cite{saccadesFriston2012a}.

In another behavioral experiment looking at embodied decision-making, Beilock and Holt found that the visual perception of stimuli can recruit the motor system into action simulation (the production of potentially detectable micro-motor outputs) which can influence affective judgments~\cite{beilockEmbodiedPreferenceJudgments2007}. Together, research within this school of thought encourages us to reconsider traditional notions of passive perception followed by active decisions, and instead see decision-making as a continuous and dynamic process. In this light, sensed biometric data may serve as a correlate to deliberative processes. The privacy implications stemming from this reasoning have not yet been sufficiently investigated.

\subsubsection{Eye Tracking, Virtual Reality, and Privacy} 

Eye-tracking technology is far from new, but the increasing prevalence of video capture systems and improved algorithms for accurately inferring gaze direction~\cite{li2005starburst} make it especially relevant to conversations around modern algorithmic surveillance. Further, the breadth of application areas being explored for eye-tracking---as a novel UI selection affordance~\cite{piumsomboon2017exploring}, a method to augment user experience via attentional awareness~\cite{vertegaal2008attentive}, and an opportunity to gain computational efficiencies via foveated rendering~\cite{patney2016towards}, among many others---reinforces the belief of some researchers that eye tracking technology will soon be ubiquitous~\cite{Liebling2014}.

Beyond eye-tracking, a critical body of literature has explored the deleterious effects of various regimes of biometric surveillance~\cite{raji2020saving,marciano2019reframing}, including works specifically highlighting the dangers when these technologies are targeted at marginalized groups~\cite{madianou2019biometric}, and the unique effects of systems purporting to monitor internal states such as affect~\cite{bullington2005affective}. The documentation of surveillance harms offered by these scholars, among others, provided a key motivation for the present work.

Scholars in HCI have offered a variety of perspectives on privacy concerns relating to the use of consumer remote sensing technologies such as Internet of things (IoT) products like smart speakers, and augmented reality (AR) devices. Denning et al.\ explored the perceptions and privacy concerns of bystanders to the use of glasses-style AR devices which may record video of people and surroundings~\cite{denning2014situ}. In a 2014 review, Roesner et al.\ identified several key challenges and risks of AR systems, including adversarial attacks to both input and output channels, and theft or misuse of sensor data~\cite{roesner2014security}. Recent work by Ahmad et al.\ raises concerns over privacy ambiguity imposed by `always-on' IoT devices, and introduces the concept of tangible privacy---design features that allow bystanders to clearly assess the state of a device's sensors~\cite{ahmad2020tangible}.

We chose to implement our study using a VR-based laboratory precisely because of the ease of collecting not only gaze targets, but a rich stream of sensor data allowing us to measure our participants' in-task behavior at a fine level of granularity. Indeed, the potential to develop machine learning-based predictive models trained on VR biometric data, some of it imperceptible to human observers, has not gone unnoticed by businesses and entrepreneurs~\cite{morrisVirtualRealityStartupStrivr2016}. The ethical questions raised by this new domain, which is quickly developing on interrelated but discrete paths in the private sector and academia, are diverse and poignant. A first of its kind conference was held in 2018 proposing to construct a ``VR bill of rights''~\cite{vrPrivacyConf}, and an opinion article by Jeremy Bailenson in the same year highlighted the capabilities and dangers of biometric data collected in VR~\cite{bailensonProtectingNonverbalData2018}.

While immersive VR clearly comes with a host of potential privacy concerns, our primary interest centers around VR as a research tool. The VR-based setup allows us to probe the sensitivity of collected data, as well as the psychology influencing the perception of surveillance, in both in-home and public settings. Ultimately, we see the virtual laboratory as offering researchers a controllable, replicable setting enabling the study of human behavior in a real world increasingly subject to surveillance.

\subsection{Research Questions}
\label{sec:rq_hypotheses}

Though biometric surveillance has attracted extensive scholarship, few quantitative behavioral studies have shed light on the expectations, assumptions, and range of behavioral responses employed by individuals hoping to evade prediction by an adversarial system. In this work, we aim to explore the following questions:

\begin{enumerate}
    \item RQ1: Under what conditions might biometric signals such as motor outputs, eye movements, and electrodermal activity expose sensitive information relating to beliefs and immediate choice intentions?
    \item RQ2: How effectively can commonly used machine learning models predict choice intention given behavioral data during decision-making?
    \item RQ3: What strategies are used by participants when instructed to make unpredictable decisions in a simple tracked setting?
    \item RQ4: How effective are the employed strategies at maintaining an individual's privacy of intent?
\end{enumerate}

We note that as per RQ2, we seek to understand the approximate performance of minimally tuned off-the-shelf tools, rather than to estimate an upper bound on prediction accuracy. The development of novel algorithms that might improve performance on this prediction task were not among the objectives of this work (see Section \ref{sec:ethics} for discussion). 

\section{Methods}

\subsection{Task Design}
\label{task_and_hypotheses}

In order to study human behaviors related to the protection of private intent, we developed a game-based experiment in which participants were sufficiently motivated to avoid predictability while behaving in a physical environment. Specifically, we defined the following desiderata to constrain experimental design:

\begin{enumerate}
    \item \textbf{Covert imperative}. Participants should be sufficiently incentivized to obscure their intent.
    \item \textbf{Non-trivial choice-salience}. Decisions made in the task must influence a relevant outcome (e.g. compensation); participants must consider an extrinsic value of their choice separate from masking intent.
    \item \textbf{Embodied}. Choices should be enacted by gross motor outputs as opposed to verbal reports to allow for action simulation or other micro-motor outputs to be detectable.
\end{enumerate}

\subsection{Covert Embodied Choice Task}

Based on these criteria, we designed a virtual reality task (inspired by the popular game \emph{Set}) as follows. 

\begin{figure}[h]
  \centering
  \includegraphics[width=\linewidth]{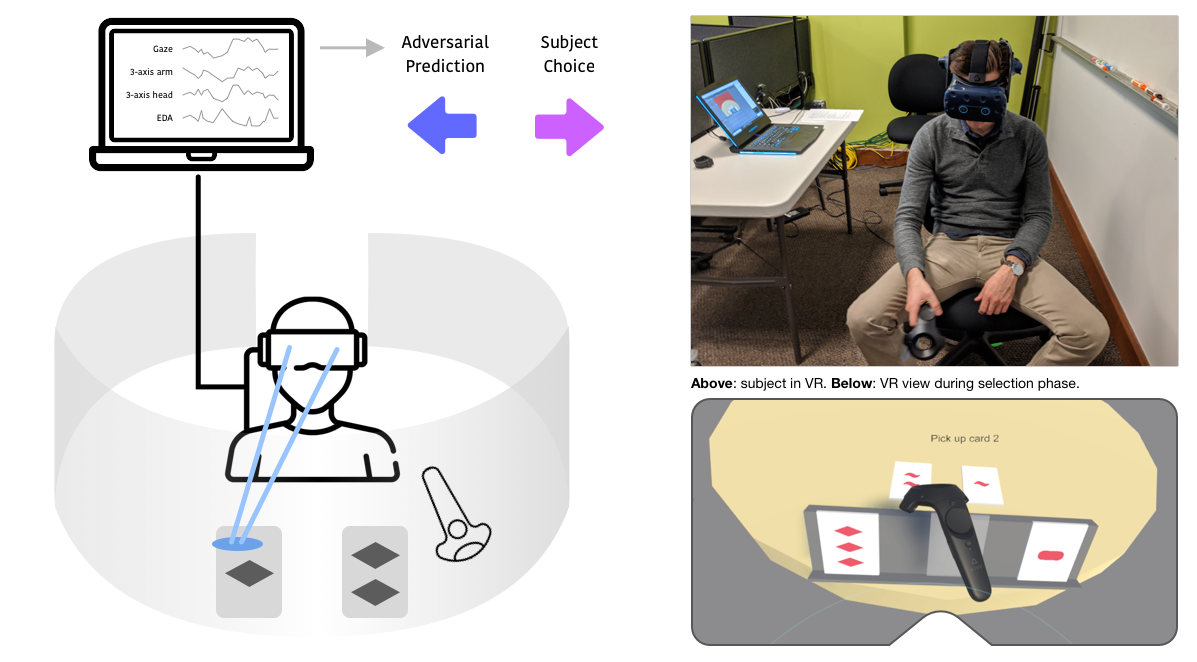}
  \captionof{figure}{Left: schematic of subject during a trial. Top-right: participant in VR. Bottom-right: participant's field of view during card selection.}
  \Description{Left: schematic of subject during a trial. Cartoon illustration of user holding VR controller, with gaze rays pointing to one of two cards on the table illustrating eye-tracking. A computer shows biometrics captured during each trial, and arrows indicate the prediction of the adversary as well as the participant's true selection. Top-right: A participant sits wearing an HMD and holding a VR controller. Bottom-right: Screen shot of participant's field of view during card selection. Controller is hovering over holder displaying two private cards, one with one diamond, one with three.}
  \label{fig:procedure}
\end{figure}

In each trial, the participant is presented with two cards in a private `hand' facing them, and two cards flat on a table (see Figure \ref{fig:procedure}). During the 10-second decision phase, they must choose one of the two single cards on the table. Then, during the 3-second selection phase, they use the controller to bring the chosen card into their hand to complete a trial. A trial is successful if the resulting three cards in the hand form a complete `match'. A successful match is defined as three cards for which both card attributes: count and shape are either all the same or all different. 
A trial is completed either upon the participant's releasing their chosen card into their hand, or when the selection timer elapses. A new trial is started after completion of the last, with a new deal of two private and two table cards. 

See Section \ref{sec:procedure} for a more detailed description of the task and experimental protocol.

\subsection{Participants}

Study participants were recruited via e-mail through an on-campus experimental lab at a public university in the United States. Participants were screened to minimize the risk of adverse effects from the use of VR: participants indicating dizziness or nausea during prior uses of VR, epilepsy, a history of seizures, or that they are prone to motion sickness, were not considered for the full study. 

To support the use of parametric tests with our mixed between-subjects and within-subjects design, we selected a target sample size of 35 subjects per group (70 total). Since machine-learning-based classification accuracy was a primary dependent variable, it was critical to ensure both that (1) the training set contained sufficient data with respect to the dimensionality of the feature-space and complexity of the model to effectively generalize, and (2) that the test set was large enough to accurately estimate model performance. Despite stopping data collection early due to the COVID-19 pandemic (with 57 completions), our resultant dataset included 1200 non-practice subject-trials which we deemed sufficient for our planned analysis.

Two participants were dropped after they indicated in the post-experiment interview that they misunderstood the instructions and believed the task required them to enable the adversary to predict their intent (PP 004), or provide training labels to the adversary (PP 041).

Participants were randomly placed into one of the two study conditions: early adversary (N=27), or delayed adversary (N=28). The two study conditions operated identically except for the trial number at which the adversary info screen appeared (see \ref{sec:procedure}), allowing us to control for potential learning effects. Participants had a mean age of 20.4 $\pm$ 4.6, 34 identified themselves as female, and 21 as male. 47.3\% reported their race as Asian, 34.5\% white, 1.8\% American Indian or Alaska Native, 1.8\% Native Hawaiian or Pacific Islander, and 14.5\% Other, including Asian \& white, and Middle Eastern. 16.4\% of participants identified as Spanish, Hispanic, or Latino. All participants were undergraduate students, graduate students or staff at the university where the study was conducted.

\subsection{Experimental Procedure}
\label{sec:procedure}

After providing consent, each participant went through a self-paced slide-based tutorial using the experiment computer. The tutorial explained the pattern matching task, and provided quizzes to ensure the participant had understood how to identify a match consistent with the rules of the game. The tutorial also detailed how a final success score would be calculated, and the monetary bonus that would be provided depending on this score. Bonuses were calculated based on the percent of successful rounds (a correct match for pre-adversary rounds, and both a correct match and the avoidance of prediction in the adversarial rounds), however information about the adversary was not presented until later. All participants received the base compensation (\$20) for their participation, with performance-based rewards contributing an additional 0-25\% of the base level. At the end of the tutorial the experimenter gave the participant an opportunity to ask any clarifying questions.

Next, the experimenter fitted the participant with an Empatica E4 skin conductance and heart rate monitor, and a Vive Pro Eye VR headset. The participant completes a brief exercise in which they fixate on targets that appear on the screen in sequence to calibrate the in-unit eye tracking system. Once complete, the participant was asked to use the trigger on the controller to begin the first of four (unscored) practice rounds in the experiment. The practice rounds simulate a normal trial, and allow the participant to become comfortable with the mechanics of the interaction in VR. Practice rounds, like regular rounds, had a 10-second decision timer, and 3-second selection timer, so that participants learned how to make their selection in the allotted time.

After completing all four practice rounds, the participants were asked to confirm they understood the task and the interaction, and if so, they used the controller trigger to begin the main experiment.

After three trials (early condition) or 22 trials (delayed condition, midway point) the participant was presented with a screen informing them that the task would be changing and that in all remaining trials, an adversary would be ``tracking their behavior during the decision phase'' and attempting to predict which card they would choose. They were also informed that success on each following trial would be determined by \emph{both} selecting a correct match \emph{and} avoiding their choice being predicted by the adversary\footnote{During experimentation, the adversary was programmed to make predictions based on a simple gaze-based heuristic trivial to compute in real-time. Because only an aggregate success score was provided to participants after all trials were complete, the specific mechanism of prediction could not influence participant behavior.}. After the participant reviewed this information, the experimenter verbally asked whether the information was clear and prompted for any final questions before instructing the participant to continue with the experiment when ready.

All participants completed all 44 non-practice trials. Upon completion of the last trial, a screen appeared summarizing the participant's performance across all trials (counts of correct matches, and successful avoidance of prediction), and displaying their final success score and bonus compensation earned. 

After removing the VR headset, the participant was asked to respond to a short questionnaire and structured interview covering their beliefs about the biometric data that was used, strategies they employed, and confidence in their efficacy\footnote{The questionnaire and interview were intended as exploratory methods to capture unexpected responses and hear from participants in their own words. While the experimental study and quantitative results are the primary contributions of this work, the qualitative data allowed for more nuanced interpretation and set up possible focus areas for future, more in-depth investigation.}. Once complete, a debrief form and media release form were signed, and finally the participant was provided their incentive payment and told they may exit the lab. 

The recruitment process and study protocol was approved by the local ethics review board.

\subsection{Data Structure, Cleaning, and Analysis}

The VR experiment was implemented in C\# using the Unity Editor (version 2019.3.0f6), Steam VR (version 1.9.16), and Steam VR Unity Plugin (version 2.5.0). Telemetry was recorded in the update loop and included 3-dimensional position and rotation vectors for the HMD (head motion), and controller (arm motion). Two types of gaze data were collected. Fixations were recorded using the Vive Pro Eye SRanipal SDK (version 1.1.0.1) which infers in-scene gaze target objects via the Tobii Gaze-to-Object-Mapping algorithm \footnote{For details on Tobii's G2OM algorithm, see \url{https://vr.tobii.com/sdk/technology/tobii-g2om/}}. This data type was converted into a series of fixation records tracking start and stop timestamps, as well as the scene element fixated. Secondly, raw gaze data was recorded in the update loop providing a gaze origin and direction vector in world coordinates, as well as an approximate vergence distance allowing tracking of the implied $(x,y,z)$ coordinates of raw gaze target.

Physiology data including heart rate, inter-beat interval (IBI), and electrodermal activity, was extracted from data files produced by the E4. We extracted EDA spike timestamps using peakutils~\cite{lucas_hermann_negri_2017_887917}, and computed heart rate variability (HRV) based on IBI records using the \emph{HRV} Python library~\cite{bartels2020hrv}. All data was aligned and registered by comparing synchronized device timestamps and trial-by-trial metadata recorded by the Unity experiment. All experiment code is available in an MIT licensed repository: \url{https://github.com/onejgordon/cec_vr}.

\subsection{Predictive Model}



To assess participants' effectiveness at protecting privacy of intent, we trained a discriminative machine learning model to predict a participant's ultimate choice of card. Trials where the participant failed to choose any card before time elapsed were excluded, so the model is required only to correctly predict a binary label corresponding to \emph{choose left} or \emph{choose right}. Data available to the predictive model included all physiological and behavioral data collected during the 10-second decision phase, which was clearly demarcated to participants by a text alert and changing background color (see section \ref{sec:feature_eng} for details). Data captured during the selection phase, such as participants' arm motion to select a card, was not available to the algorithm. 

We conducted experiments within two related prediction paradigms both consistent with plausible real-world settings but presenting unique challenges.

In the first, we trained a \textbf{participant agnostic (PA) model} to predict participant choice without any identifying information about the participant or their behavior in other trials. The prediction problem is posed in the classical machine-learning format, to learn a predictive model: $ M_{PA} = P(Y_t|X_t) $, where $Y_t$ is the choice label for trial $t$, and $X_t$ is the corresponding feature vector.

In the second paradigm, we trained an ensemble of \textbf{behavioral-typology (BT) models} that first group participants based on several rounds of trial data (including choice labels). The model is then required to predict participant choice for all remaining trials. This paradigm was motivated by the qualitative observation that participants' choices often became predictable to the experimenter once a strategic pattern was detected. 

As such, the behavioral-typology model is given trial data and labels from the first 3 adversarial trials completed by each participant. The prediction problem for each trial $t$ becomes:

$ M_{BT} = P(Y_t|X_t,X_{0,1,2},Y_{0, 1, 2}) $

Here, $Y_t$ and $X_t$ are defined as before, and $X_{0,1,2}$ and $Y_{0,1,2}$ are the concatenated data and labels respectively from the first three trials.

To test this model, we chose a simple heuristic based on the most choice-informative feature according to exploratory analysis of the training set: the proportion of eye fixations on the chosen card. Participants were assigned a behavioral-typology based on this metric ($< 33\%$, $< 34-66\%$, and $> 67\%$). Independent classifiers were trained for each of the three typologies of training participants. At test time, the fixation heuristic was again used to assign a typology to each test participant (based on an estimate using only the first 3 adversarial trials), and inference was performed with the associated classifier. 

\subsubsection{Training Paradigm}

Both prediction problems used the same dataset which was produced by randomly assigning 50\% of participants into each set ($N_{train} = 28$ and $N_{test} = 27$), and then building a feature matrix (see following Section \ref{sec:feature_eng}) where each row corresponded to a single non-practice trial (excluding those without a successful card selection) from a single participant, with \emph{set} (train vs. test) determined by the participant's assignment. As such, all models were tested using trials from novel participants, with no opportunity for train-test leakage. The resultant matrix contained 1200 trials of which 52\% were used for training. To compare predictive performance and the success of obfuscation, independent models were trained on pre- and post-adversarial trials separately.

We tried a number of off-the-shelf classification algorithms for this task, and report results for two that performed best overall: Scikit-learn's implementation of the Random Forest Decision Tree (RFDT) and Gradient Boosted Decision Tree (GBDT) classifiers~\cite{scikit-learn}. Due to the limited sample size and lack of a separate evaluation set, only minimal hyper-parameter tuning was performed in order to avoid overfitting the test sample.

To assess each model's performance we report accuracy scores (percent of trials correctly predicted) across all test participants' trials. 

\subsubsection{Feature Engineering}
\label{sec:feature_eng}

\paragraph{Fixations and Gaze}

To conform with the training paradigm which called for the generation of a single binary prediction per trial, we employed a number of common techniques to extract lower-dimensional feature vectors from the high-dimensional raw gaze and fixation data collected during each trial's decision phase. 

Fixations, which were tallied with a start and stop timestamp and target object, were used to produce features indicating the fixation count ($F_i$) and fixation fraction ($\frac{F_i}{\sum_j{F_j}}$) for each key object in the scene (e.g. left card, right card, holder, table). Since each table card was expected to be a particularly informative region of interest, we also computed the minimum, maximum, and mean duration of fixations on each. Additional fixation-based features included: last and second to last fixation object, and percent of trial for which eyes were closed.

In contrast to discrete fixations, features based on raw gaze points have the potential to capture behavior in which participants look near but avoid direct fixation upon an object of interest. We computed simple descriptive features from the mean, mode, and standard deviation of each of the three coordinate dimensions.

\paragraph{Head and Controller Motion}

To capture dynamics of the controller and head mounted display (HMD), which measures head position and rotation, we included descriptive statistics from the position and rotation of each (e.g. HMD Y-position standard deviation, controller yaw mean, final HMD roll). Position and rotation features included an absolute version, as well as one relative to trial start. We estimated mean and maximum velocity based on position deltas and timestamps.

Finally, to capture more granular information that might indicate the trajectory of head and arm motion, we broke the decision phase into 10 equal duration segments, and included features for 3-axis position snapshots at each time point.

\section{Results}
\label{sec:results}

Our analysis indicates significant changes when comparing behavior between adversarial and non-adversarial trials, as well as substantial individual differences during performance of the task. In Section \ref{sec:strategy} we review results, including the post-task questionnaires and interviews, related to apparent behavioral strategies and avoidance of prediction.

\subsection{Effect of Adversary}
\label{sec:ae}

In the following sections we report results of analyses comparing data before and after each participant received the adversarial instruction. Unless stated, analyses include participants from both early adversary and delayed adversary conditions. The physiological and behavioral variables analyzed exist on a vague continuum from signals typically considered involuntary (e.g. heart rate variability and electrodermal activity), to those usually interpreted as the result of intentional, conscious action (e.g. gross arm movements). The very premise of this study rests on the ambiguous middle ground, where participants may attempt to control or alter behaviors they know to be under observation in order to obfuscate their true intent. Which hand motions, head tilts, or saccades are the product of voluntary control versus involuntary physiological or motor outputs cannot be established by our analyses, or indeed any algorithm~\cite{juarrero2000dynamics}. Though post-task interviews do shed light on the strategies consciously selected by participants, self-reports provide only a subset of all effects on behavior exhibited. For the purposes of this study, however, we need not discriminate voluntary versus involuntary action, and rather ask only how measurable behaviors change, and to what extent these changes affect the ability of an algorithmic agent to successfully predict a participant's ultimate choice. For an analysis of behavior changes consistent with self-reported strategies, see Section \ref{sec:strategy}.

\subsubsection{Physiology}
\label{sec:ae_phys}

\begin{figure}
\centering
    \begin{subfigure}[t]{\linewidth}
        \centering
        \includegraphics[width=\linewidth]{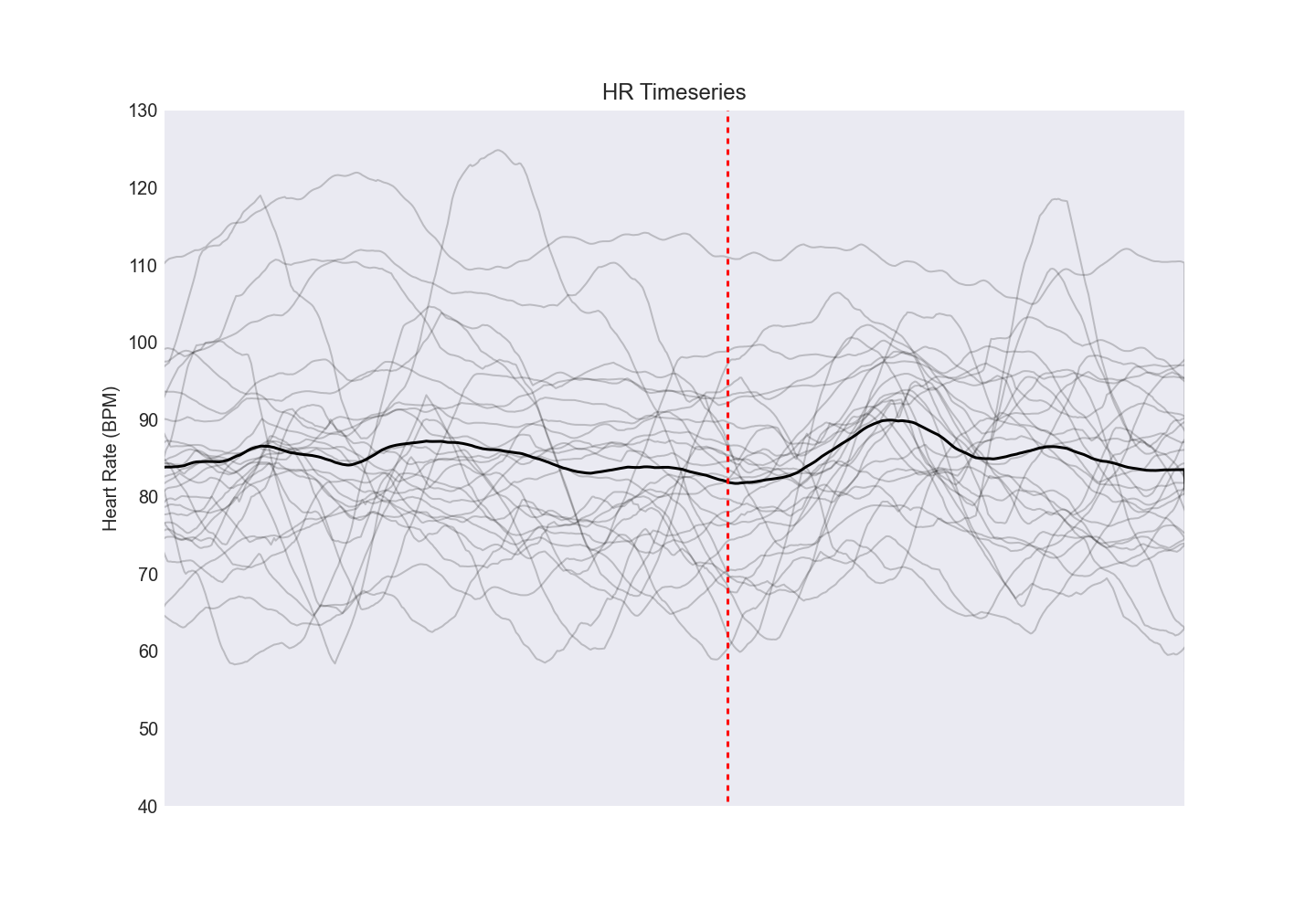}
        \caption{Heart rate time-series for all delayed condition participants, across full session duration. Red dashed line indicates adversarial transition.}
        \Description{Timeseries for each participant's heart rate across entire duration of session. Mean heart rate is shown in bold, and shows a roughly flat trend with a slight peak after the advent of the adversary, which is indicated with a vertical red dashed line. }
        \label{fig:hr_phys}
    \end{subfigure}
     \hfill
     \begin{subfigure}[t]{\linewidth}
        \centering
        \includegraphics[width=\linewidth]{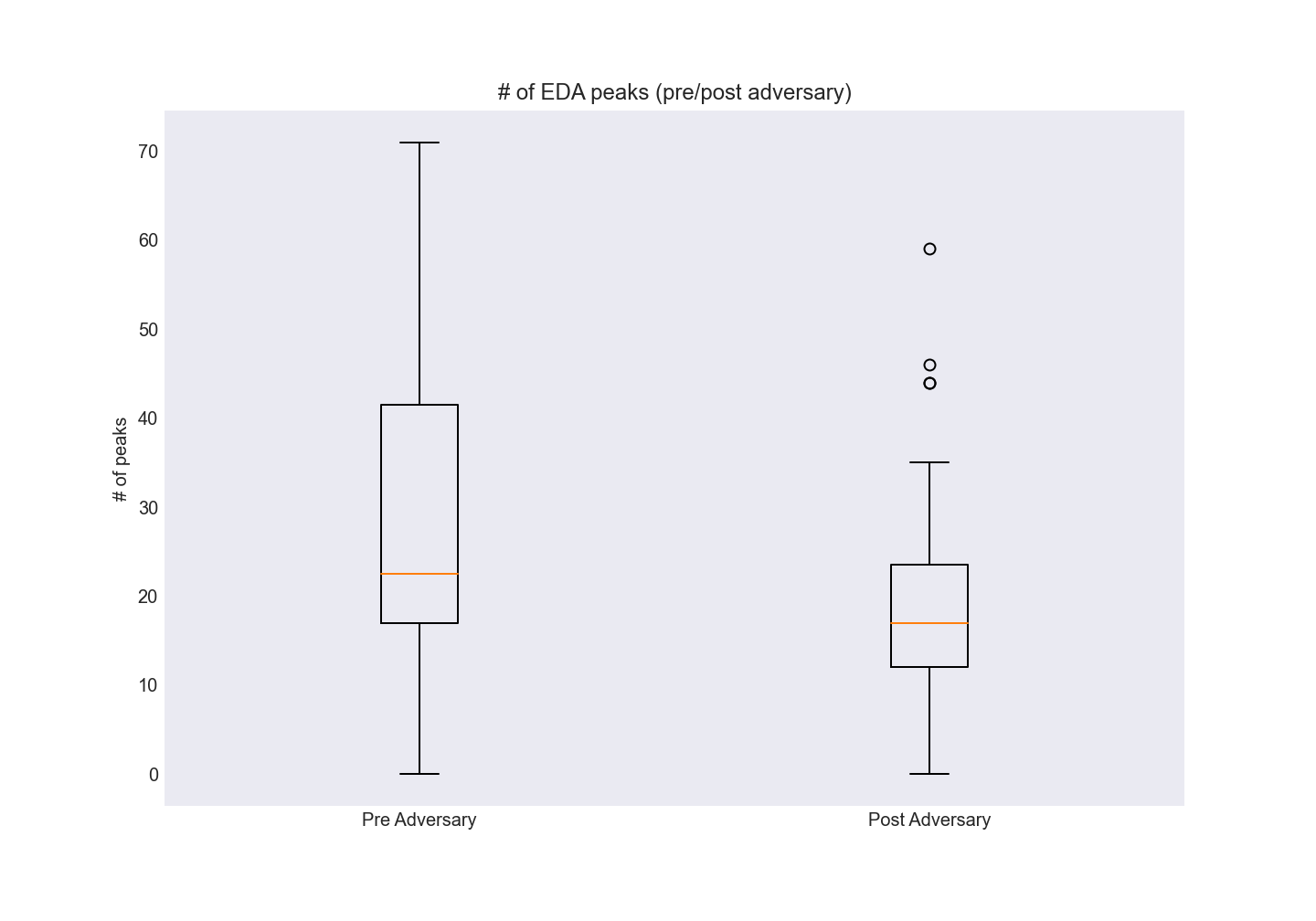}
        \caption{Count of EDA spikes, pre- and post-adversary. Reduction in spikes was not statistically significant according to independent $t$-test.}
        \Description{Box plot comparing EDA spikes pre- and post-adversary. Post-adversary mean is slightly lower, but ranges of box and whiskers overlap.}
        \label{fig:eda_phys}
    \end{subfigure}
    \caption{Analysis of physiology data before and after adversary.}
\end{figure}

Though a slight peak is observed after adversarial trials start, we found no statistically significant difference in heart rate between non-adversarial and adversarial trials. Results suggest that the number of EDA spikes reduced post adversary ($t=1.92$, $p=0.059$), though temporal and learning effects are additional potentially confounding causes (see Figure \ref{fig:hr_phys} and \ref{fig:eda_phys}).

\subsubsection{Movement}
\label{sec:ae_move}

\begin{figure}
\centering
    \includegraphics[width=\linewidth]{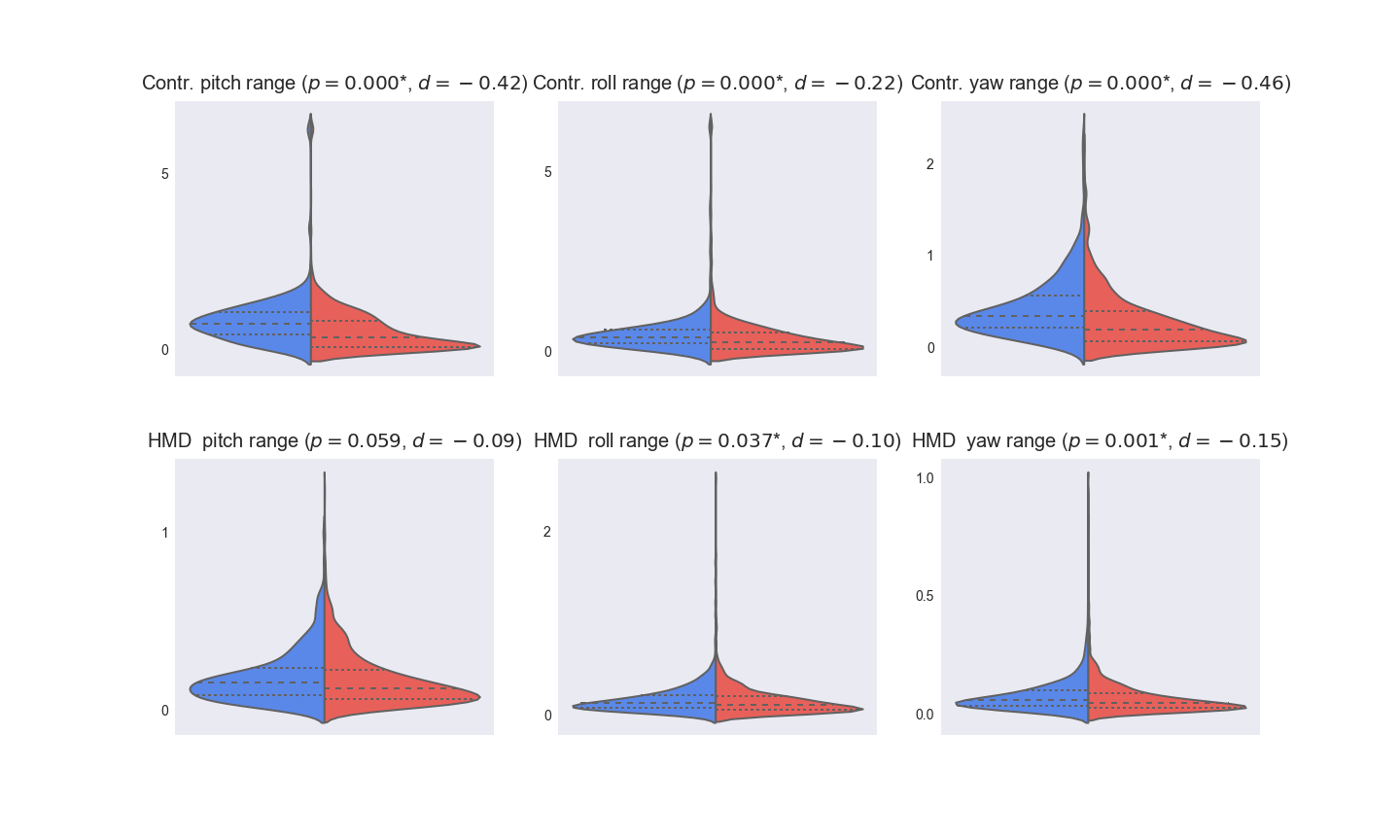}
    \caption{Head and controller range pre- (blue) and post-adversary (red). Range for both reduced significantly post-adversary suggesting support for signal dampening strategy. $p$-value for independent sample $t$-test, and Cohen's $d$ shown above each plot. }
    \Description{Six violin plots indicate the change in HMD and controller pitch/roll and yaw range pre- and post-adversary. p-values and Cohen's d in the title of each plot show that all differences but HMD pitch are statistically significant at the p<0.05 level.}
    \label{fig:hmd_ctrl_range}
\end{figure}

In adversarial trials, participants tended to move their arm and head less. Specifically, we assessed the range of motion along each rotational dimension (pitch, roll, yaw), as $\theta_{range} = max(\theta) - min(\theta)$. Range of all three dimensions for both controller (arm) and HMD (head) was significantly lower in post-adversary trials (statistical test results shown in Figure \ref{fig:hmd_ctrl_range}). 

\subsubsection{Gaze}
\label{sec:ae_gaze}

The distribution of raw gaze data also significantly shifted during adversarial rounds. Both the mean and standard deviation of gaze target height ($y_{mean}, t: -3.55, p<0.001$; $y_{SD}, t=-3.55, p<0.001$) were significantly lower with adversary.

\subsection{Strategic Behavior and Prediction}
\label{sec:strategy}

\subsubsection{Self-Reported Strategies}

Post-task questionnaires and interviews probed participants on how they interpreted the requirements of the post-adversary task, how they modified their behavior (if at all) to avoid prediction, as well as the experience of being asked to make a decision under adversarial surveillance. While the primary motivation for these interviews was to verify participants' comprehension of the task, they contain some useful insights into conscious strategies used, as well as the challenges posed by the task overall.

\begin{figure}[h]
\centering
    \includegraphics[width=\linewidth]{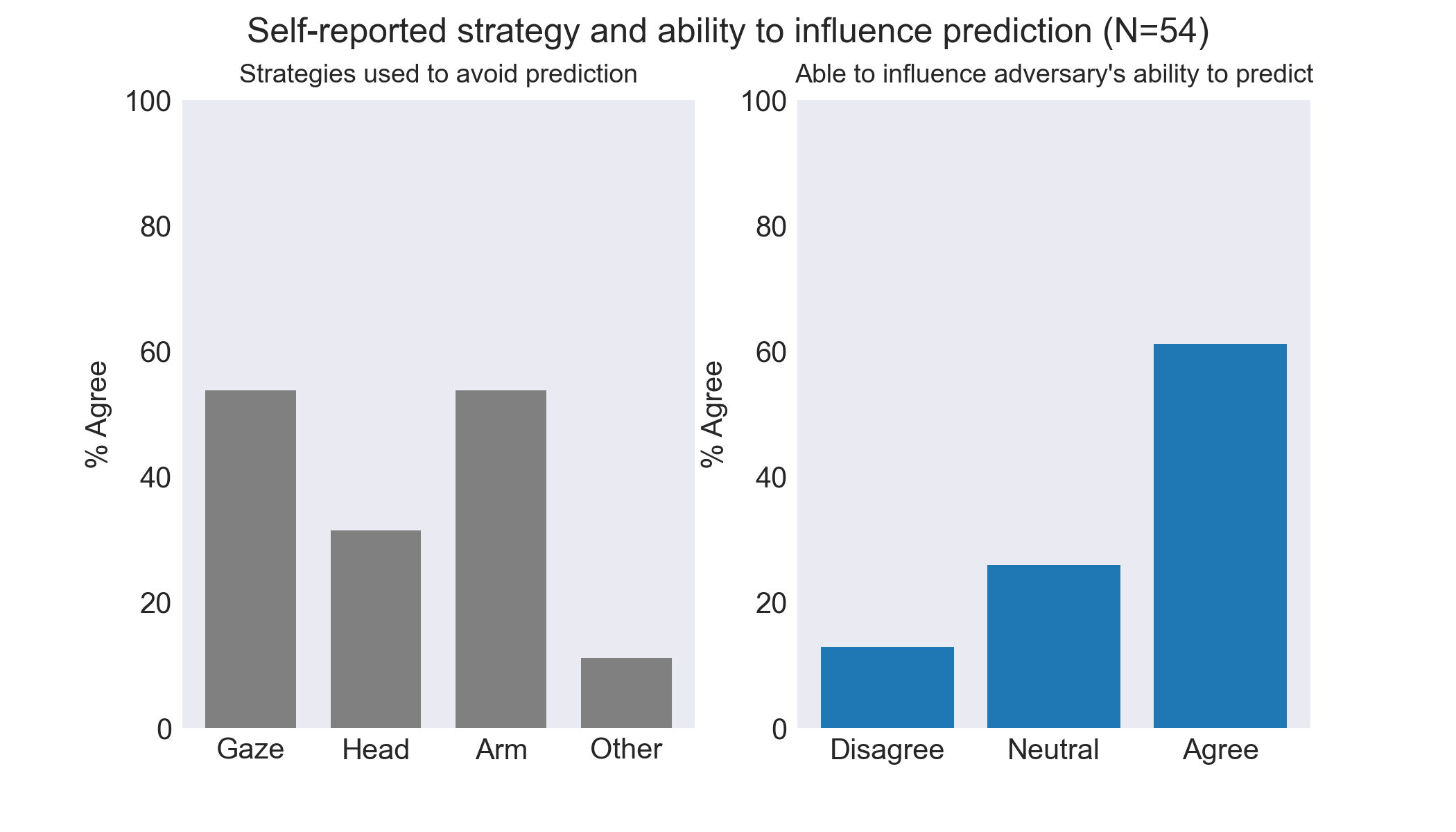}
    \caption{Results from post-task survey. Left: Participants reported modifying multiple behaviors with gaze and arm motion most often selected. Right: 61\% of participants responded 'Agree' or 'Strongly Agree' when asked if they believed they were able to successfully influence the adversary's ability to predict their choice.}
    \Description{Two plots with title "Self-reported strategy and ability to influence prediction (N=54). Left: bar plot showing strategies used to avoid prediction. Gaze and Arm are selected by over half of participants. Right: bar plot of responses to agreement that participants could influence adversary's ability to predict. The agree bucket is largest with over 60\% of participants.}
    \label{fig:self_report_strategies}
\end{figure}

Questionnaire responses indicate that participants primarily adjusted behavior related to gaze and arm motion (see Figure \ref{fig:self_report_strategies}). In interviews, participants identify a range of meta-strategies which we categorize as signal dampening, signal randomizing and overt misdirection. Signal dampening strategies included the suppression of behaviors that participants believed may give away their intent, such as holding their head still, reducing motion in their arm, pointing their hand in a neutral direction, entraining gaze on a neutral portion of the table or on the timer, and looking near but not directly at the cards on the table. Signal randomizing techniques included both adding excess behaviors during each trial, such as shifting gaze widely and continuously, moving their head back and forth at random, as well as randomizing overt asymmetrical behaviors by pointing the controller sometimes at the chosen card, and sometimes away. 

Overt misdirection included intentional actions that `telegraph' intent opposite to true choice, such as fixating gaze or head position for long durations on the opposite card, moving the controller towards the opposite card during the decision phase, etc.

Several participants reported initially looking at or near the table cards just long enough to make their decision, then switching to a misdirection or obfuscation strategy for the remainder of the decision phase: ``I would look at both and figure out what the right answer was, and then I would stare at the wrong one, or go off into my own thoughts'' (PP 012). Other participants talked about an evolving thought process on obfuscation, and the realization that consistent misdirection may also be predictable: ``At first I thought, oh it makes sense to try not to do what you would normally do, like if you're going to pick this, then look at the other one for longer. After a while I was like, oh that's also predictable, so maybe switch it up.'' (PP 015).
  
Over 60\% of participants believed they were able to influence the adversary's ability to predict their intent using the strategies they employed. 

\subsubsection{Agency and Cognitive Load}

While the majority of participants reported that they believed they were able to influence the adversary's prediction, some weren't sure, and others disagreed. In post-task interviews, some of these participants noted that they didn't understand how the adversary was making its predictions, and so felt it wasn't clear how they could effectively respond (PP 046, PP 016, PP 057): ``I guess I just didn't really know what it was going to be looking at, so I didn't know what to change'' (PP 017). Others found the matching aspect of the task challenging, and felt they needed to focus on selecting the right card rather than avoiding prediction: ``I was more preoccupied with getting the right card, so kind of forgot about [the adversary] sometimes'' (PP 003). For another, the decision and the mechanics of making their selection took priority: ``I was still trying to pick the right card, like grab it in the right amount of time'' (PP 016).

These assessments highlight that for some participants uncertainty about the dynamics of algorithmic prediction undermined their feeling of agency. Additionally, the cognitive load imposed by the combination of the card matching task and time limit were challenging for some participants to balance with the need to obfuscate. In the following section we review results comparing biometric data when left is chosen versus when right is chosen, thus identifying choice-correlated asymmetries likely exploited by the predictive model.

\subsubsection{Correlates of Choice}

To detect behavioral patterns that might be leveraged by a predictive model, we performed a variety of exploratory analyses comparing raw data and features collected during the decision phase, segmented by participant choice. Any asymmetries seen either across the full training set, or within a substantial sub-group, might indicate strategic behaviors or involuntary predictors correlated with intent. We report several such asymmetries below.  

\paragraph{Gaze}

\begin{figure}
\centering
    \includegraphics[width=\linewidth]{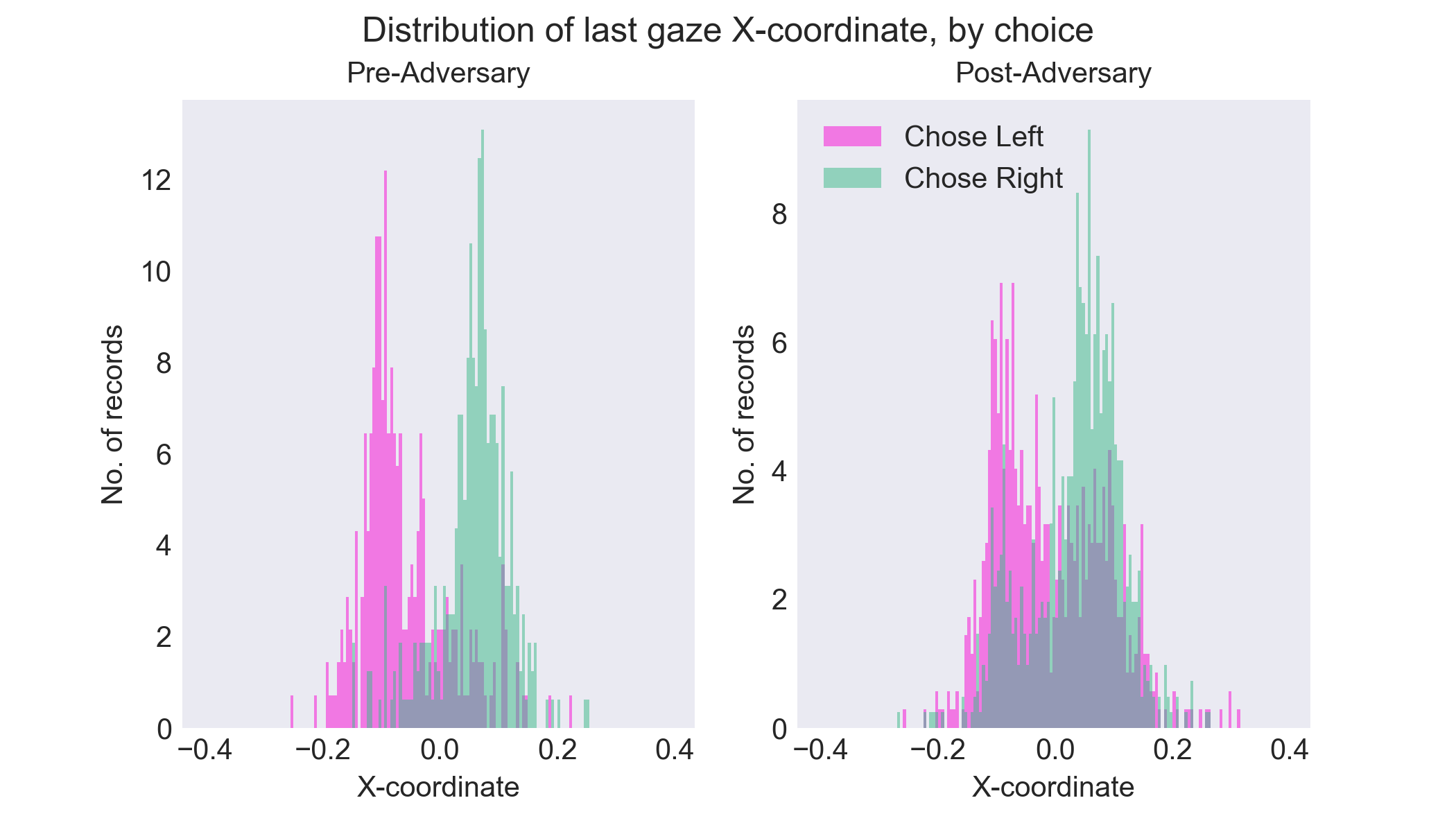}
    \caption{The increasing overlap of the gaze x-coordinate distribution between left vs. right trials post-adversary illustrates behavior change to deter prediction based on gaze. On average, however, final gaze point was highly predictive of choice both pre- and post-adversary.}
    \Description{Two distribution plots side by side, with main title "Distribution of last gaze X-coordinate, by choice". Left: distribution of X-coordinate for pre-adversary trials. Right: distribution of X-coordinate for post-adversary trials. Both distributions show significant discrimination, but right plot exhibits more overlap.}
    \label{fig:raw_gaze_x}
\end{figure}

Even when averaging across the full training population, the x-coordinate of the final raw gaze point was highly predictive of choice ($ d=-0.447, t=-8.68, p<0.0001$, see Figure \ref{fig:raw_gaze_x}).

\paragraph{Longitudinal Observations}
\label{par:longitudinal}

Though comparing summary statistics of gaze and motion data highlights systematic differences in behavior, this approach ignores the richness of temporal visual attention inherent to this decision-making paradigm. By analyzing individual participants movement and gaze longitudinally (both across trials, and within each trial), additional behavioral regularities appear. 

\begin{figure}[h]
\centering
    \includegraphics[width=\linewidth]{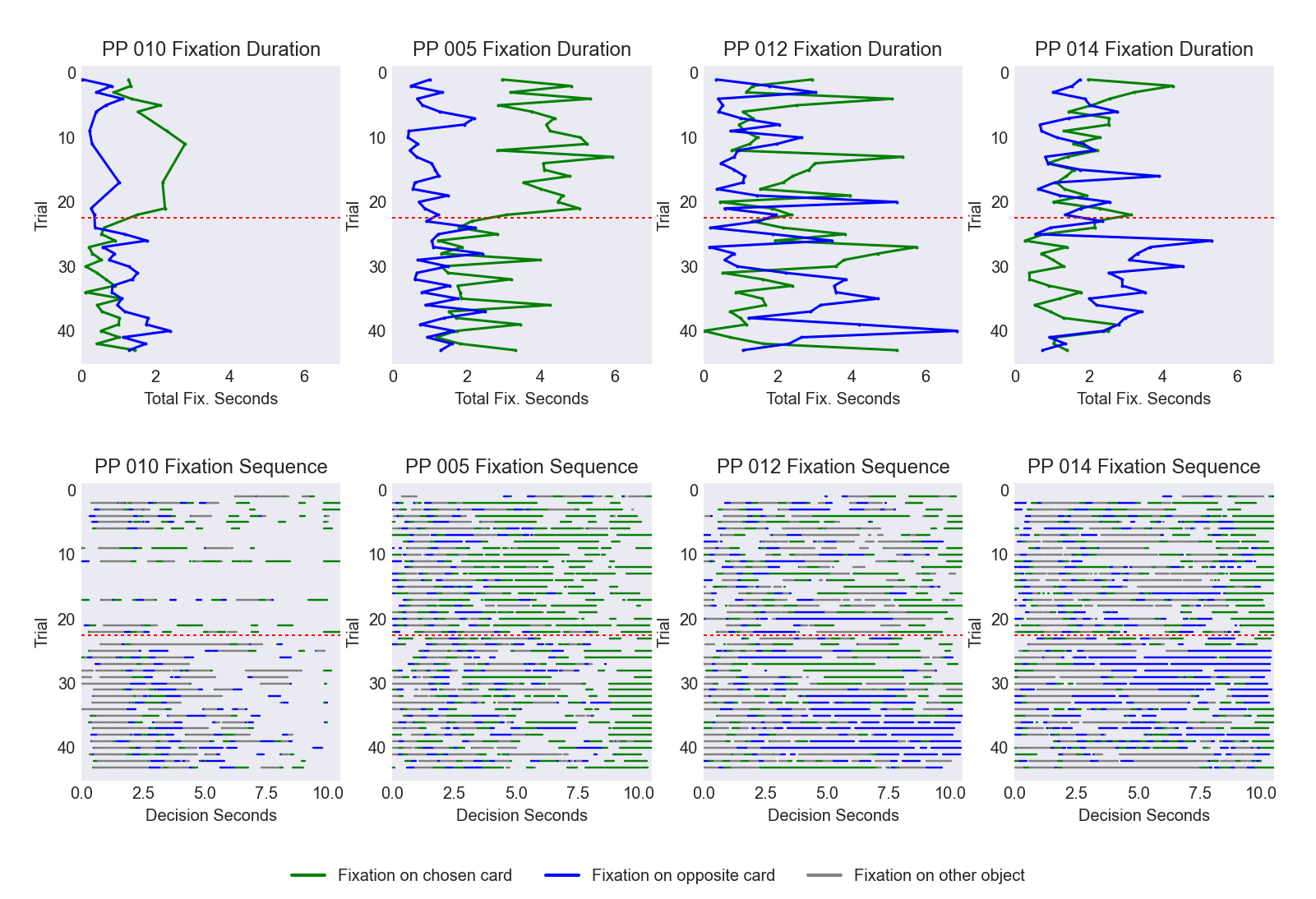}
    \caption{Total fixation duration (top) and fixation time-series (bottom) by trial for 4 selected participants (PPs). Green: fixations on chosen card; Blue: fixations on other card; Gray: fixations on non-card objects (shown in sequence charts only). Trials progress vertically from top to bottom, and the dotted red line indicates the beginning of adversary trials. Note that trials for which no card was selected are omitted.}
    \Description{Eight plots illustrate fixation dynamics across trials for four selected participants. The top row plots duration of fixation on each table card (chosen and opposite), while the bottom row shows fixation sequences during the 10 second decision phase. See caption and body text for discussion.}
    \label{fig:fix_timeseries_4}
\end{figure}

Figure \ref{fig:fix_timeseries_4} illustrates trial-wise fixation time-series and total duration by card (chosen versus not chosen) for selected individual participants. These plots demonstrate dynamics of fixations, and behavioral change post-adversary, found to be representative of common behavioral patterns among the training population. 

Participants PP 010 and PP 005 both favored fixations on the card they eventually chose prior to adversary, but show very different behavior post-adversary. Post-adversary, PP 005 reduces fixation duration on their chosen card, but both fixation duration, and final fixation (note the regularity in final fixation) still indicates choice on nearly every post-adversary trial. PP 010, on the other hand, is likely using a misdirection strategy and intentionally switches visual attention to the opposite card.

PP 014's fixation behavior illustrate a key finding---in some cases, conscious obfuscation may unintentionally unmask intent. While fixations are well-balanced pre-adversary, the participant's choice becomes significantly more predictable post-adversary under an apparent strategy of fixating on the opposite card.

Overall, these analyses highlight the value of providing temporal gaze sequence features (in addition to duration and fixation count) to the predictive models.

\paragraph{Arm Movement}

\begin{figure}
\centering
    \includegraphics[width=\linewidth]{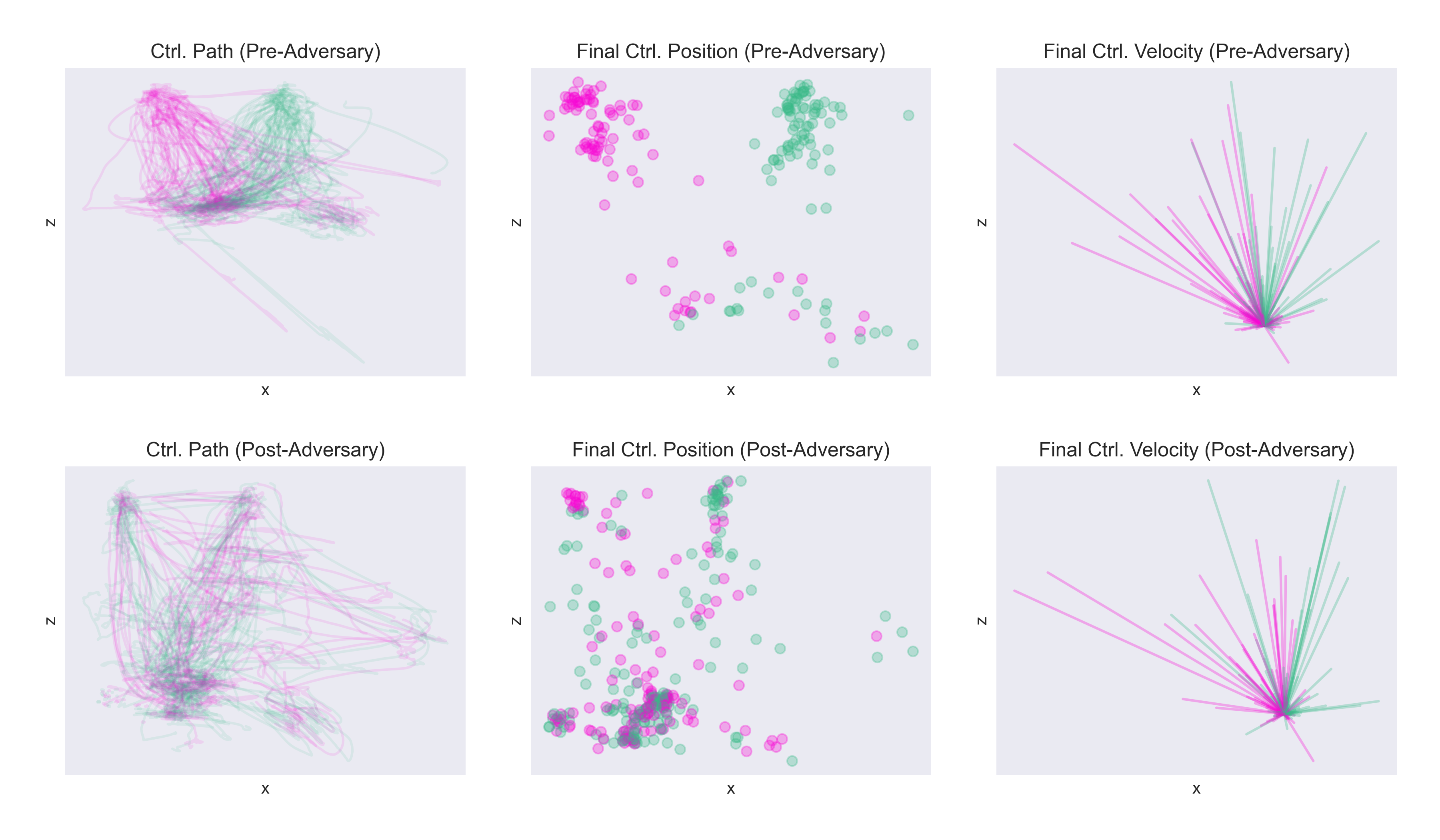}
    \caption{Top-down view of controller trajectories (left), final position (middle) and final velocity (right), at termination of decision phase. Turquoise: right selected; Magenta: left selected. Clear choice correlation can be seen in all three metrics pre-adversary. Post-adversary data exhibits less stereotyped segmentation, but clear differentiation is still visible in both final position and final velocity.}
    \Description{Six plots illustrate a top-down view of controller path, final controller position, and final controller velocity (as vectors from a shared origin). The top row shows each plot pre-adversary, and the bottom shows each plot post-adversary. See body text and caption for discussion.}
    \label{fig:ctrl_trajectories}
\end{figure}

An analysis of arm movement (measured by controller trajectories), reveals a substantial behavior change pre- and post-adversary. Trajectories are consistently stereotyped by choice in early non-adversarial trials, and show many participants preemptively (prior to selection phase) moving their controller over the chosen card in preparation for selection. In post-adversary trials, we see support for self-reported strategies involving constrained controller motion as well as misdirection (e.g. the density of final controller positions over the opposite card in the bottom center plot in Figure \ref{fig:ctrl_trajectories}).

Despite this visible obfuscation behavior, choice-correlation remains in final controller position, and particularly in final velocity. We discuss these findings further in Section \ref{sec:discussion}.

\subsubsection{Algorithmic Prediction}

\paragraph{Participant-Agnostic Predictor}

\begin{figure}
\centering
    \includegraphics[width=\linewidth]{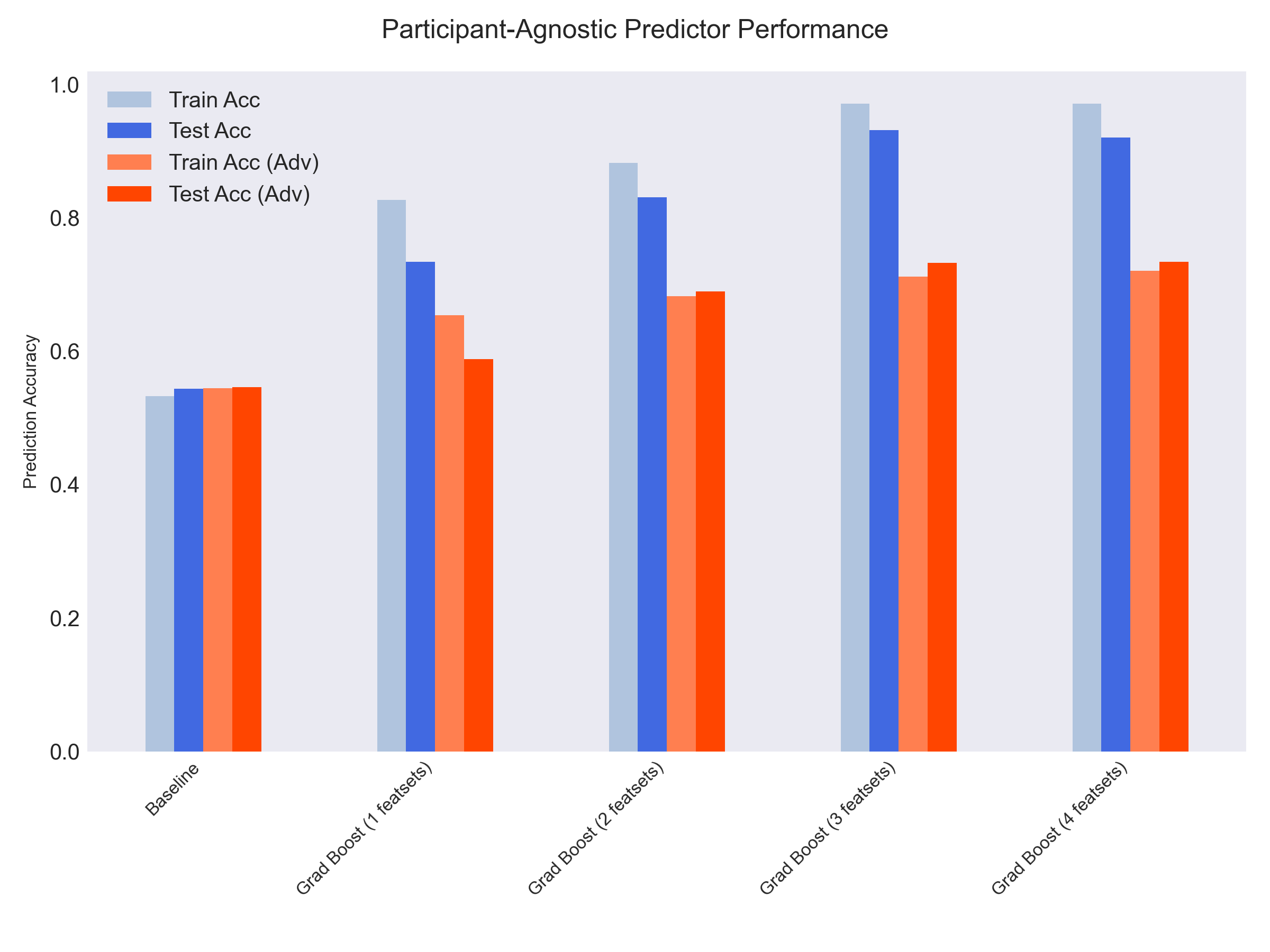}
    \caption{Model accuracy comparison for participant-agnostic predictor, with forward addition of feature-sets. Feature sets were added as follows: 1) Gaze features, 2) Fixation features, 3) HMD features, 4) Controller features. 
    }
    \Description{Single plot with main title "Participant-Agnostic Predictor Performance". Bar chart shows forward addition of feature-sets and train and test accuracy for each, comparing pre-adversary and post-adversary performance. Bars increase and reach a maximum around 90\% accuracy (pre-adversary) and around 70\% (post-adversary).}
    \label{fig:model_comparison_sa}
\end{figure}

Even in the more challenging participant-agnostic paradigm where the predictive model must identify features successfully correlating with choice without the context of prior participant behavior, prediction rates were high. The best-performing model was the Gradient Boosted Decision Tree (GBDT), with a prediction accuracy of 93\% on pre-adversary trials, and 73\% on post-adversary trials (see Figure \ref{fig:model_comparison_sa}). 

The most predictive features included last gaze fixation target, mean gaze X-coordinate, and final HMD roll offset. 

\paragraph{Behavioral-Typology Predictor}

\begin{figure}
\centering
    \includegraphics[width=\linewidth]{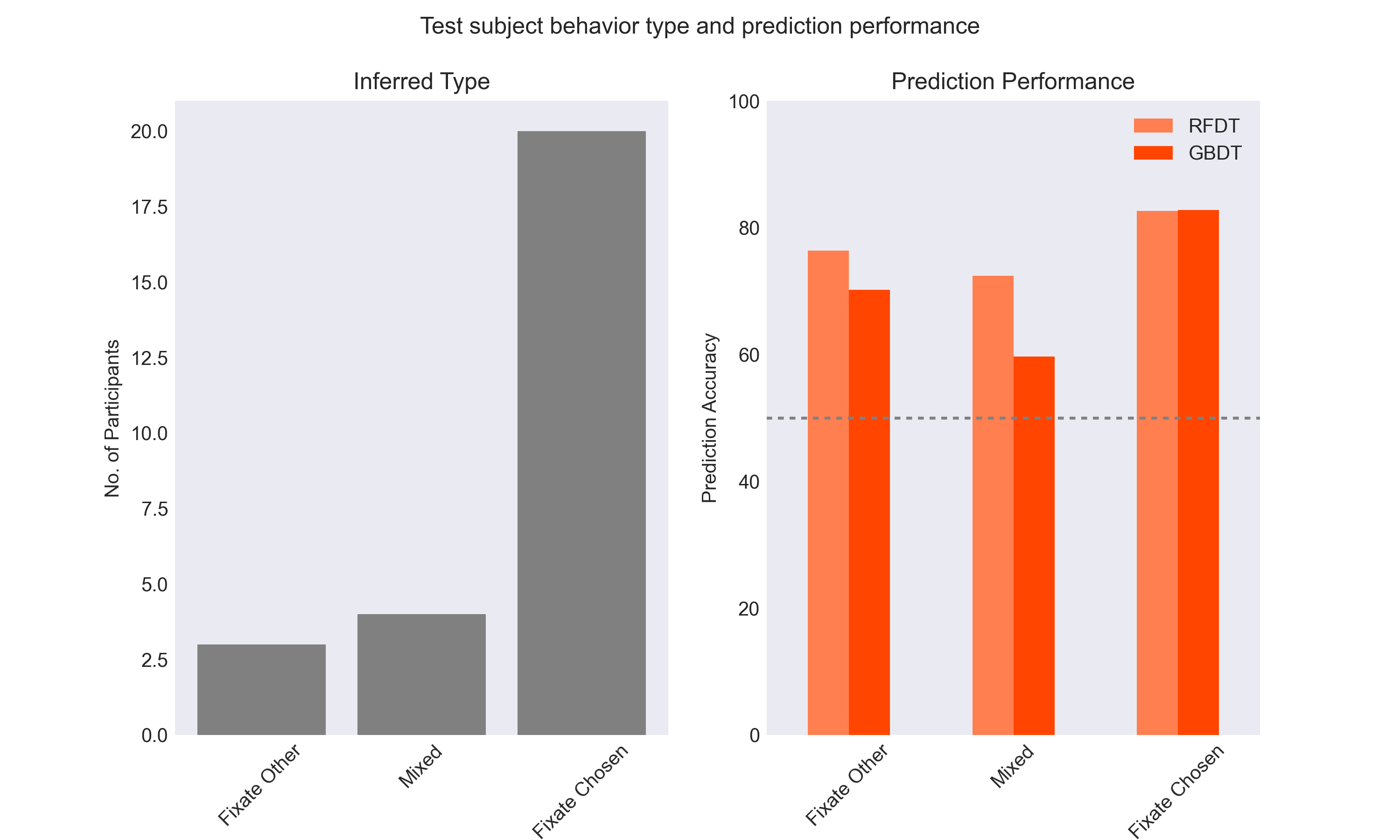}
    \caption{Behavioral-Typology model performance for each inferred typology based on proportion of fixations: fixations prefer chosen card, fixations prefer other card, and mixed.}
    \Description{Two plots side by side with main title "Test subject behavior type and prediction performance". Left: a histogram of inferred type with the majority of participants falling into "Fixate Chosen". Right: a bar chart titled "Prediction Performance" showing each model's prediction accuracy for each typology. The highest accuracy is for "Fixate Chosen" but all three typologies demonstrate accuracy with the RFDT predictor above 70\%. }
    \label{fig:seg_performance}
\end{figure}

As anticipated, the behavioral-typology (BT) predictor improved overall prediction accuracy, achieving 72.4\% for the most effective (least predictable) strategy typologies, and 82.7\% for the least effective behavioral typologies (see Figure \ref{fig:seg_performance} for BT model performance details).

\begin{figure}[h]
\centering
    \includegraphics[width=\linewidth]{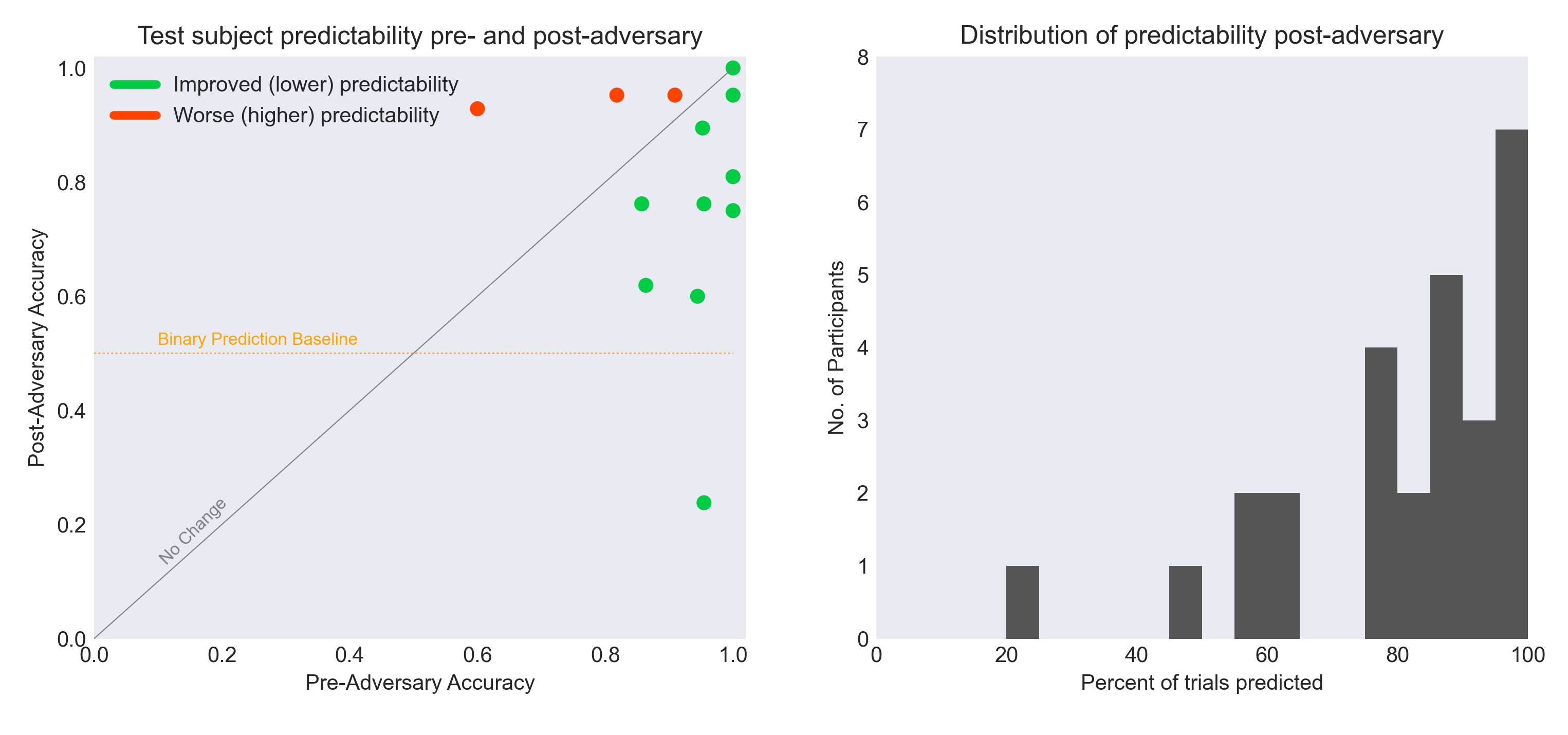}
    \caption{BT model predictability pre-adversary (x-axis) compared to post-adversary (y-axis). Some test participants (red) became more predictable post-adversary. The distribution of predictability (right) illustrates that the majority of participants' choices could be successfully predicted in between 75\% to 100\% of their trials. Note: the chart at left includes only test participants in the delayed adversary condition, since BT model performance cannot be assessed from the three pre-adversary trials. The chart at right includes all test participants across both conditions.}
    \Description{Two plots side by side. Left: A scatter plot titled "Test subject predictability pre- and post-adversary" shows a cluster of participant points in the upper right corner. A diagonal line splits the plot into two triangles. Most points fall within the lower triangle and are colored green, three points appear in the upper triangle and are colored red. Right: a histogram titled "Distribution of predictability post-adversary" shows the number of participants falling into each "Percent of trials predicted" bin. More than half of participants fall in the bins above 75\% accuracy.}
    \label{fig:predictability_change}
\end{figure}

The majority of participants are reliably predicted in between 75 to 100\% of trials using this technique (see accuracy distribution post-adversary in Figure \ref{fig:predictability_change}).

Table \ref{tab:results} provides a summary of model prediction accuracy.

\paragraph{Strategic Efficacy}

We can quantify the efficacy of a participant's change in behavior by comparing our trained model's prediction performance between pre- and post-adversary trials. We define a metric for strategic efficacy ($\Psi$) as follows:

$$\Psi = \frac{N_{correct,pre}}{N_{total,pre}} - \frac{N_{correct,post}}{N_{total,post}} $$

If efforts to conceal intent are effective, prediction accuracy post-adversary will be lower (positive $\Psi$). Figure \ref{fig:predictability_change} plots BT model accuracy pre- and post-adversary thus illustrating strategic efficacy (distance from diagonal). Participants for whom prediction accuracy decreased are plotted in green, while increases in prediction accuracy are plotted in red. The clustering of test participants below the diagonal illustrates that, through strategic adjustment to behavior, most participants were able to successfully reduce choice prediction, but not by much. Indeed, only one participant (shown at lower right) evaded prediction in more than half of trials, and on average across participants, 80.4\% of post-adversary trials were successfully predicted.

Also visible in Figure \ref{fig:predictability_change}, is the surprising finding that for a number of participants (red dots), efforts to become less predictable had the opposite effect producing an increased prediction accuracy post-adversary.

To further investigate behavioral and physiological differences that might explain variance in strategic efficacy, we compared two groups of test participants: those less than 70\% predictable post-adversary, which we term `effective', and those at least 70\% predictable post-adversary, which we term `ineffective'. The latter group can be seen as the rightmost disjoint cluster in the histogram in Figure \ref{fig:predictability_change}. We find that the effective group exhibited statistically significant reduction in controller movement (yaw std. dev.: $d=1.544, t=3.012, p = 0.006$; final pitch: $d=0.932, t=9.950, p < .001$; roll range: $d=0.583, t=5.469, p < .001$), head movement (pitch std. dev.: $d=0.745, t=6.964, p < .001$; roll range: $d=0.848, t=8.111, p < .001$; yaw range: $d=0.692, t=6.758, p < .001$), percent of fixations on chosen card ($d=1.470, t=2.588, p = 0.016$) and total fixation duration on chosen card ($d=1.263, t=2.205, p = 0.037$). These results together indicate that, on average, the effective group moved their head and arm less during the decision phase, while also fixating less frequently and for shorter durations on the card they ultimately selected.

\begin{table}
  \caption{Summary of predictive model performance. Best-performing models for each prediction paradigm are listed for both pre-adversary (pre) and post-adversary (post) trials, Gradient Boosted Decision Tree (GBDT) for the Participant-Agnostic (PA) predictor, and Random Forest Decision Tree (RFDT) for the Behavioral-Typology (BT) predictor.}
  \label{tab:results}
  \begin{tabular}{ccc}
    \toprule
    Model & Train Accuracy & Test Accuracy\\
    \midrule
    GBDT (PA, Pre) & 97.1\% & 92.0\% \\
    GBDT (PA, Post) & 72.0\% & 73.4\% \\
    RFDT (BT, Pre) & -- & 91.0\% \\
    RFDT (BT, Post) & -- & 80.4\% \\
  \bottomrule
\end{tabular}
\end{table}

\section{Discussion}
\label{sec:discussion}

\subsection{Behavior and Prediction}

The experimental paradigm presented here was designed to provide insights into both the sensitivity of behavioral and biometric data when leveraged towards the prediction of future decisions (RQ1/2), the response of individuals confronted with a specific kind of adversarial surveillance (RQ3), as well as the interaction between the two in the effect of awareness of tracking on an individual's ability to maintain private intentions (RQ4).

Classification accuracy results indicate that collected data were extremely predictive of choice (over 90\% accuracy) for participants during pre-adversary trials. Though pre-adversary prediction was performed primarily to produce a baseline against which to compare performance during adversarial trials, this finding alone deserves some consideration. In the absence of any reason to mask intent, behavioral data collected during deliberation was sufficient to predict eventual choice in the vast majority of trials.

In adversarial trials, after being informed of the imperative to decide which card to select without revealing intent, most participants adopted strategies based on logical assumptions about the dynamics of prediction. Participant interviews revealed a thought process involving the various data they believed might be used by the adversary. The gaze calibration exercise at the beginning of the VR session suggested to many participants that their eye movements may be among the data used by the adversary to anticipate their decision. Seeing their controller in VR was another strong signal that arm motion might be a target of tracking, as well as an opportunity for masking or obfuscation. Some participants observed that during pre-adversary trials their hand would begin to move just before the end of the decision phase, and reported attempting to adjust, invert, or randomize this behavior post-adversary. Despite correctly identifying these signals, however, components of gaze, head, and arm motion all remained correlated with choice when averaged across the test sample. These correlations, in aggregate, allowed the classifiers within both prediction paradigms to successfully predict card choice in spite of the strategies employed. 

To understand why, it is instructive to revisit one of the most informative data types captured: gaze fixations. In Figure \ref{fig:fix_timeseries_4}, PP 010 and PP 005 both demonstrate significant adjustments to gaze dynamics upon the beginning of adversarial trials. Both strategies, once identified, remain predictive of choice---PP 005 reduces overall fixations on either card, and appears to randomize fixations while deciding, but fixates their chosen card at the end of each decision phase (green lines at end of post-adversary trials). PP 010 also reduces overall card fixations, and inverts their gaze behavior in a way that remains highly correlated with choice.

Optimal behavior, as any child fond of rock-paper-scissors (or game theory) intuitively knows, depends upon the randomization of behavior in order to minimize correlations between observable data and intended action. Randomization, even of a single series of discrete choices, however, is known to be impossible for humans to achieve~\cite{rapoport1992generation}. In contrast to univariate randomization, our task is especially challenging due to its multi-dimensionality---it requires successful randomization of all detectable behaviors simultaneously.

Finally, several of the strategies detected appear consistent with the types of intuitive behavior that might be effective at masking intent from a human observer. This is one interpretation of results showing lowered head pitch (Figure \ref{fig:hmd_ctrl_range}), and lowered gaze, which fail to obfuscate the target of gaze from eye-tracking hardware like that used in this study.

\subsection{Limitations and Future Work}

The prediction accuracy results achieved in our analysis are not meant to establish a ceiling. Rather than developing an optimal behavior prediction tool, the aim of this work was to assess the feasibility of intent prediction, as well as the range of individuals' responses to an explicitly surveilled task paradigm. With a larger training set, it is likely that even the standard statistical techniques used in this analysis would learn an improved behavioral model capable of exploiting the regularities of a wider variety of strategies and therefore achieve prediction accuracy improvements. 

An additional limitation stems from the simplistic behavioral-typology inference model which we based on a single metric (card fixation ratio) indicative of a single behavioral strategy. With a slightly larger participant pool, an unsupervised learning approach identifying natural subject clusters or axes of behavioral variation would likely have produced improved prediction accuracy for the BT model. 

While our results suggest dynamics that may extend into real-world contexts in which individuals interact with surveillance systems, the typical considerations of the generalizability of in-lab findings apply here. While a significant performance-based incentive was used to motivate task success, behavior in a setting with more significant potential consequences---as is increasingly relevant to real-world surveillance---might well display dynamics different from those observed in this study. Relatedly, though the game-based setting was conducive to our analysis, studying participants as they make more naturalistic decisions (e.g. who to vote for, or how much to tip) may offer insights that more easily extend to everyday behavior.

We hope future work will build on this preliminary study to develop a more holistic understanding of individuals' reasoning about and response to biometric surveillance.

\subsection{Implications}

The kinds of privacy risk supported by our findings are likely best addressed through a combination of public awareness, regulatory action, and the concerted efforts of designers~\cite{ardagna2009obfuscation, toubiana2011trackmenot, pierce2019smart}.

Among other proposed solutions, Bailenson suggests users may take it upon themselves to use hardware filters capable of adding noise and reducing the fidelity of collected data~\cite{bailensonProtectingNonverbalData2018}. Our findings, however, suggest that users may overestimate the efficacy of more intuitive obfuscation strategies, and underestimate the sensitivity of the data collected from them. As such, hoping to encourage costly user action such as obtaining and using obfuscation devices may not be a reliable solution.

Public education, however, must certainly play a role. Alphabet's ``Digital Transparency in the Public Realm'' initiative is one example that hopes to encourage cities and other actors collecting data in public spaces to use a set of colored privacy icons to indicate the types and sensitivity of data being collected~\cite{YourDataBeing}. The behaviors and misconceptions illuminated by our results, however, indicate some critical limitations of this effort as well. What will an individual who notices an eye-tracking symbol on a light-post do with this information? Will Facebook require users of its recently announced AR glasses to wear warning labels to inform passersby of ongoing tracking~\cite{AnnouncingProjectAria2020}? What use is informing members of the public about surveillance, without instructions for opting-out? And if, as was the case for some of our participants, strategies to protect one's privacy instead act to exposes sensitive information, might messaging like this cause additional harms?

Even more fundamentally, however, any response to tracking technologies must contend with the fact that what is being collected is a moving target: the signals hidden within raw data evolve with new algorithms, new users, and integrations with additional data sources. The sensitivity emerges not from the data itself, but what can be done with it. This study highlights the counterintuitive nature of ``what can be done,'' and as such the challenges privacy advocates must overcome when communicating these ideas to the public.

\subsection{Ethical Considerations}
\label{sec:ethics}

It is crucial, when conducting work either employing or studying technologies that have been used to harm individuals and communities, to critically exam the potential impact on these populations and society as a whole. We reason that this research is unlikely to reinforce harms, and holds the potential to bring valuable data to academic and public dialogue regarding the capabilities, risks, and vulnerabilities that may be exploited by algorithmic surveillance.

First, this work does not develop any novel algorithm or computational model that might improve predictive performance, but rather explores the efficacy of off-the-shelf machine learning tools. Second, that the findings of this work likely parallel research conducted privately within organizations seeking to leverage algorithmic prediction in the interest of financial or political gain. If our speculation related to potential prediction accuracy is correct, we must acknowledge the multiple orders of magnitude that separate the sample size and scope of data used in this study with that available to companies and other organizations already in the business of collecting biometric data from individuals.

\section{Conclusion}

In this work, we examined the expectations individuals hold about biometric surveillance, and how these beliefs influence behavioral response to a tracked setting. Our results suggest that participants hold a range of priors about the nature of biosignals that might be leveraged for prediction, and use a wide variety of strategies to attempt to make a future choice less predictable. While some participants questioned their agency in evading the adversary, most modified their behavior and successfully reduced their prediction accuracy. However, data collected remained highly predictive of choice (over 80\% mean accuracy), and the majority of participants were correctly predicted by the behavioral-typology model in 75-100\% of trials. Importantly, a meaningful subset of participants adopted a strategy that on average increased the model's ability to successfully predict their choice, suggesting the counterintuitive nature of the dynamics of algorithmic prediction.

\section{Acknowledgments}

This material is based upon work supported by the National Science Foundation Graduate Research Fellowship Program under Grant No. 2019236659. Any opinions, findings, and conclusions or recommendations expressed in this material are those of the author(s) and do not necessarily reflect the views of the National Science Foundation.

\bibliographystyle{ACM-Reference-Format}
\bibliography{bibliography}

\end{document}